\documentclass[pre,10pt,onecolumn,notitlepage,superscriptaddress]{revtex4-1}
\usepackage{amsmath,amsthm}   
\usepackage{graphicx}   
\usepackage{subfigure}  
\usepackage{hyperref}  
\usepackage{amsmath}
\usepackage{amssymb}
\usepackage{mathtools}
\usepackage{multirow}

\usepackage{tikz,pgfplots}
\usetikzlibrary{pgfplots.groupplots}

\newcommand{\sumtwo}{\sum_{\nn\in \mathbb{Z}^2\setminus\{\mathbf 0\}}}
\newcommand{\sumd}{\sum_{\nn\in \mathbb{Z}^d\setminus\{\mathbf 0\}}}
\newcommand{\defeq}{\equiv}
\newcommand{\dhr}{\delta\hat\rho}
\newcommand{\E}{\mathcal  E}
\newcommand{\n}{\mathcal{N}}

\newcommand{\rr}{\mathbf r}
\newcommand{\bb}{\mathbf b}
\newcommand{\nn}{\mathbf n}
\newcommand{\blue}{\mathcal B}
\newcommand{\red}{\mathcal R}

\newcommand{\R}{\mathbb{R}}

\newcommand{\J}{\mathrm{J}}
\newcommand{\bnd}{\beta^{(2)}_N(d)}
\newcommand{\bn}{\beta^{(2)}_N}
\newcommand{\bnu}{\beta^{(p)}_N}          
\DeclareMathOperator{\dd}{d}
\newcommand{\abs}[1]{{\left|#1\right|}}
\newcommand{\norm}[1]{{\left\|#1\right\|}}    	
\DeclareMathOperator{\grad}{grad}
\DeclareMathOperator{\re}{Re}
\newcommand{\Hess}{\mathsf{Hess}} 
\DeclareMathOperator{\dive}{div}   
\newcommand{\Z}{\mathbb{Z}}
\DeclareMathOperator{\e}{e}    

\newcommand{\ebmp}{\textsc{Ebmp}} 
\newcommand{\emmp}{\textsc{Emmp}}

\begin{document}
\title{Scaling hypothesis for the Euclidean bipartite matching problem}
\author{S. Caracciolo}
\affiliation{Dipartimento di Fisica, Universit\`a degli Studi di Milano and INFN,
via Celoria, I-20133 Milano, Italy}
\author{C. Lucibello}
\affiliation{Dipartimento di Fisica, Universit\`a ``La Sapienza'', P.le A. Moro 2, I-00185, Rome, Italy}
\author{G. Parisi}
\affiliation{Dipartimento di Fisica, INFN -- Sezione di Roma1, CNR-IPCF UOS Roma Kerberos, Universit\`a ``La Sapienza'', P.le A. Moro 2, I-00185, Rome, Italy}
\author{G. Sicuro}
\affiliation{Dipartimento di Fisica ``Enrico Fermi'', Universit\`a di Pisa, and INFN -- Sezione di Pisa, Largo Bruno Pontecorvo 3, I-56127, Pisa, Italy}

\begin{abstract}
We propose a simple yet very predictive form, based on a Poisson's equation, for the functional dependence of the cost from the density of points in the Euclidean bipartite matching problem. This leads, for quadratic costs,  to the analytic prediction of the large $N$ limit of the average cost in dimension $d=1,2$ and of the subleading correction in higher dimension. A non-trivial scaling exponent, $\gamma_d=\frac{d-2}{d}$,  which differs from the monopartite's one,  is found for the subleading correction. We argue that the same scaling holds true for a generic cost exponent in dimension $d>2$.
\end{abstract}
\newpage
\maketitle
\section{Introduction and Main Results}
\label{empSection}
The \textit{matching problem} is a combinatorial optimization problem that has drawn the attention of both the computer science \cite{Kuhn,Munkres1957,papadimitriou1998combinatorial,Edmonds1965,Bayati2008} and the statistical physics \cite{Mezard1985,Mezard1986a,Martin2001a,Parisi2002} communities for many decades. Even in its most general formulation, the problem belongs to the $P$ computational complexity class, and many famous algorithms have been developed to solve it efficiently \cite{Kuhn,Munkres1957}.

In this paper we focus on a restricted version of the problem, the \textit{complete bipartite matching problem}, also known as the \textit{assignment problem}, that, by itself, has a long standing tradition among the scholars \cite{Mezard1985,papadimitriou1998combinatorial,mezard2009information}: in this problem we have two sets of the same cardinality $N$ (let us call them $\mathcal R$ and $\mathcal B$) and we want to find a one-to-one correspondence between the elements of $\mathcal R$ and the elements of $\mathcal B$ in such a way that all the elements are paired and a certain global function of this \textit{matching} (called the  \textit{cost function}) is minimized. An instance of the assignment problem is a $N\times N$ matrix $w$: each element $w_{ij}$ gives the partial cost of the assignment of the element $i\in\mathcal B$ to the element $j\in\mathcal R$. From a combinatorial point of view, an assignment is a permutation $\pi\in\mathcal S_N$, where $\mathcal S_N$ is the set of permutations of $N$ elements. Its cost is defined by
\begin{equation}\label{ocost}
E_N[\pi;w]=\frac{1}{N}\sum_{i=1}^{N} w_{i\pi(i)}.
\end{equation}
The optimization problem consists in finding the \textit{optimal assignment} $\pi^*$, i.e., the assignment $\pi^*$ that satisfies the property $E_N[\pi^*,w]=\min_{\pi\in\mathcal S_N}E_N[\pi;w]$.

Some interesting questions arise when random instances of the problem are considered, that is, when the elements of the cost matrix $w$ are  chosen according to a certain probability law. We will discuss the properties of the system for different choices of the disorder, i.e., of the distribution of $w$. Let us consider the average optimal cost $E_N=\overline{E_N[\pi^*;w]}$,
where $\overline{\,\bullet\,}$ denotes the expectation over the instances $w$ of the problem. If we choose the problem ensemble such that the matrix elements $w_{ij}$ are i.i.d. random variables, we obtain the so-called \textit{random assignment problem}. This version of the problem was largely investigated in a set of papers in which both the distribution of the optimal weights in the large $N$ limit \cite{Mezard1985,Aldous2001} and the finite sizes corrections \citep{Parisi2002} were derived using different approaches. In this context, the celebrated replica approach, directly borrowed from the theory of spin glasses and disordered systems, proved to be an excellent tool to investigate the properties of these random optimization problems, and led to the celebrated formula  $\lim_{N\to\infty}E_N =\frac{\pi^2}{6}$ under certain assumptions over the  distribution of $w_{ij}$ \cite{Mezard1985}.
Quite interestingly the average optimal cost $E_N$ for the random assignment problem is known exactly for every value of $N$. It is given by the simple formula $E_N=\sum_{k=1}^N \frac{1}{k^2}$, as has been first conjectured in \cite{Parisi1998} and independently proven  in Refs. \cite{Linusson2004} and \cite{Nair2003}.

From an analytical point of view, a much more difficult problem arises when the elements of the cost matrix $w$ are correlated. This is indeed the case of the \textit{Euclidean assignment problem}, also known as \textit{Euclidean bipartite matching problem} (\textsc{Ebmp}).
The \textsc{Ebmp} is an assignment problem in which a set of $N$ ``blue'' points $\mathcal B=\{\mathbf b_i\}_{i=1}^N$ and a set of $N$ ``red'' points $\mathcal R=\{\mathbf r_i\}_{i=1}^N$ are given on the hypercube $[0,1]^d$. Each point is supposed to be generated independently and uniformly at random in  the hypercube. Periodic boundary conditions are imposed (in other words, to avoid scaling corrections due to border effects, we consider the sets of points on the torus $\mathsf T^d\defeq\R^d/\Z^d$). The cost of the matching between two points is then given by a function of their distance on the torus. We will generalize the Euclidean flat distance on the torus to a family of functions characterized by a cost exponent $p$, assuming that
\begin{equation}
w_{ij}=\norm{\mathbf b_i-\mathbf r_j}^p,
\label{wnub}
\end{equation}
where $\norm{\mathbf b_i-\mathbf r_j}\coloneqq\sqrt{\sum_{k=1}^d\left(\min\left\{\abs{b^k_i-r_j^k},1-\abs{b^k_i-r_j^k}\right\}\right)^2}$ is the Euclidean norm on the torus. Due to the underlying Euclidean structure, the elements of $w$ present very strong correlations. 
We shall denote with $E^{(p)}_N(d)$ the average cost of the optimal assignment between $N$ red points and $N$ blue points on $\mathsf T^d$ with cost exponent $p$, that is 
\begin{equation}\label{optcost}
E^{(p)}_N(d)\defeq\overline{E^{(p)}_N[\pi^*;\{\mathbf r_i,\mathbf b_i\}]},
\end{equation}
where the average is intended over the positions of the points and $\pi^*$ is the optimal permutation. In the following we will sometimes drop the dependence on $p$ and on $d$ of the optimal cost, and write simply $E_N$.
 
The scaling behaviour of the leading order of the optimal cost is well known for $p>1$ and all values of $d$ and has been confirmed also by the investigations conducted by means of statistical physics methods. In fact, from a simple heuristic argument \cite{Mezard1988}, we expect that, given a red point, the nearest blue points can be found approximately in a volume of order $O\left(\frac{1}{N}\right)$ around it: their distances from the red point is, for this reason, of order $N^{-\frac{1}{d}}$. Supposing that each red point is matched to one of its nearest blue points, the expected total cost scales as $E_N=O\left(N^{-\frac{p}{d}}\right)$ for large $N$. It turns out that this asymptotic estimation  is correct only for $d\geq 3$ as it can be rigorously proved   \cite{Talagrand1992}. In a fundamental paper on this subject, published in 1984,  \citet{Ajtai1984} proved that for $d=2$ a logarithmic correction appears, $E_N=O\left( \left(\frac{\ln N}{N}\right)^{\frac{p}{2}}\right) $.  In  dimension $d=1$ instead, the divergence from the expected result is even greater, in fact it is informally known to the literature \cite{Ajtai1984,Boniolo2014} (even if to our knowledge nowhere formally stated), that $E_N=O\left(N^{-\frac{p}{2}}\right)$ in this case. We can resume the state of the art of knowledge regarding  the asymptotic behaviour of the average optimal cost in the \textsc{Ebmp}, with the following formula:
\begin{equation}
\beta^{(p)}_N(d)\equiv \frac{E^{(p)}_N(d)}{N^{-\frac{p}{d}}}=
\begin{cases}
O(N^\frac{p}{2}) & d=1,\\
O\left((\log N)^\frac{p}{2}\right) & d=2,\\
 e_d^{(p)}+O(N^{-\gamma_d}) & d>2.
\end{cases}
\label{scenario}
\end{equation}
The determination of the exponents $\gamma_d$ is one of the contributions of the present paper and it will be discussed in the following. The coefficients $e_d^{(p)}$ are not known to the literature and could not be derived  using our approach either, even though we give accurate numerical estimates in Table \ref{tab:nu1}. The rescaled average optimal cost, $\beta^{(p)}_N(d)$, is what a physicist would call the intensive energy, and will be used in the rest of the paper. As with $E_N$, we will sometimes drop the $p$ and $d$ dependence from it and write simply $\beta_N$.

The scenario depicted in Eq.~\eqref{scenario} has to be compared with the one arising in the \textit{Euclidean monopartite matching problem} (\textsc{Emmp}).  \citet{Mezard1988} studied analytically this  problem considering the correlations among the costs as perturbations around the case with random independent entries $w_{ij}$. In the \textsc{Emmp} there is a unique set of $2N$ points to be matched among themselves. It has been proven \cite{steele1997probability,yukich1998probability} that, in the \textsc{Emmp}, the rescaled average optimal cost $\beta^{(p)}_N(d)$ has a finite limit and is a self-averaging quantity in every dimension $d$. The odd behaviour noticed in the bipartite case  in low dimensions is due to the presence of differences, in small regions of space, between the number of red and blue points, that imply the presence of ``long distance'' pairings and the failure of arguments based on subadditivity \cite{Houdayer1998}. Obviously, in the monoportite cases such problems do not exist, since a partial matching between the points in an arbitrary subregion of $[0,1]^d$ leaves only one point at most unpaired.

Moreover numeric and analytic arguments  \cite{Houdayer1998} show that in the $\emmp$ the first subleading correction to the large $N$ limit of $\beta_N$ is of order $O(N^{-1})$ in any dimension. This assumptions though was also been improperly used in the $\ebmp$ to numerically extrapolate the value  $e_d^{(p)}=\lim_{N\to\infty}\beta^{(p)}_N(d)$ for $d>2$ in the case of flat distances \cite{Houdayer1998} (i.e. $p=1$). This led to some inaccurate estimations of $e_d^{(1)}$ that we address in Table \ref{tab:nu1}.
In fact in Section \ref{sec:generic} we give numerical evidence that the appropriate value for the exponent $\gamma_d$ of the subleading correction to $\beta_N$ in dimension $d>2$, as defined in Eq.~\eqref{scenario}, is
\begin{equation}
\gamma_d =\frac{d-2}{d},
\label{gammad}
\end{equation}
for any value of $p$. Notice that in the mean-field limit $d\to\infty$ one recovers the subleading scaling $O(N^{-1})$ of the random assignment problem  \cite{Mezard1986a,Parisi2002}.

The main focus of this work is the $\ebmp$ with quadratic costs, i.e. the case $p=2$ in Eq.~\eqref{wnub}. In Section \ref{sec:nu2}, inspired by some considerations on the continuum equivalent of the matching problem, the so called Monge--Kantorovi\v{c} problem, we present a powerful ansatz, Eq.~\eqref{ansatz}, for the asymptotic dependence of the optimal cost from the realization of the disorder. After a careful treatment of the diverging quantities, through an appropriate renormalization procedure, we obtain a whole new set of analytic predictions for $\beta_N$. In fact we recover the whole scenario given in Eq.~\eqref{scenario} for $p=2$, deriving the proposed expression Eq.~\eqref{gammad} for  $\gamma_d$ as well. Moreover we refine the classification given in Eq.~\eqref{scenario}, with
\begin{equation}
\beta^{(2)}_N(d)\sim 
\begin{cases}
\frac{1}{6} N + e^{(2)}_1 & d=1,\\
\frac{1}{2 \pi}\ln N + e^{(2)}_2 & d=2,\\
 e^{(2)}_d  +\frac{\zeta_d(1)}{2\pi^2} N^{-\gamma_d}  & d>2,
\end{cases}
\label{contribution}
\end{equation}
where $\zeta_d(x)$ is the Epstein zeta function. Here and in the following the symbol $\sim$ means  that the term on the $l.h.s.$ is asymptotically equal to the $r.h.s.$ except for some additional term decaying faster than each term in the $r.h.s.$ (e.g. $\beta^{(p)}_N(1)=\frac{1}{6} N + e^{(2)}_1 +o(1)$). While the coefficients $e^{(2)}_d$ have to be determined numerically, we managed to obtain analytically the coefficients of the leading order expansion of $\beta_N$ for $d=1,2$ and of the subleading order for $d>2$. In the following Sections we give a detailed derivation of these results.

\section{A scaling hypothesis for the quadratic cost}
\label{sec:nu2}

\subsection{The Monge--Kantorovi\v{c} problem and Monge--Amp\`ere Equation}
\label{sec:monge}
Let us now briefly introduce the so called Monge--Kantorovi\v{c} problem, as the conclusions of this paragraph have a crucial role in the following discussion. Given two measure densities $\rho_1\colon \mathsf{T}^d\to\R^+$ and $\rho_2\colon \mathsf{T}^d\to\R^+$, $\mathsf{T}^d=\R^d/\Z^d$ being the $d$-dimensional flat torus, $\int_{\mathsf T^d}\rho_1(\mathbf x)\dd^dx=\int_{\mathsf T^d}\rho_2(\mathbf x)\dd^dx=1$, we define  $\mathcal M$ as the  set of measure preserving maps $\boldsymbol\mu\colon {\mathsf T^d}\to{\mathsf T^d}$, i.e. the set of all maps $\boldsymbol\mu$ such that:
\begin{equation}
\rho_1(\mathbf x)=\rho_2(\boldsymbol\mu(\mathbf x))\det 
\J_{\boldsymbol\mu}(\mathbf x)\qquad\forall\mathbf x\in{\mathsf T^d}\ ,
\label{h}
\end{equation}
where $\J_{\boldsymbol\mu}(\mathbf x)$ is the Jacobian matrix of $\boldsymbol\mu$, $\left(\J_{\boldsymbol\mu}(\mathbf x)\right)_{ij}\defeq\frac{\partial\mu_i}{\partial x_j}(\mathbf x)$. Given a transportation cost function $w\colon{\mathsf T^d}\times{\mathsf T^d}\to\R^+$, we introduce the cost functional
\begin{equation}
\mathsf{E}[\boldsymbol\mu;w]=\int_{{\mathsf T^d}}w(\mathbf x,\boldsymbol\mu(\mathbf x))\rho_1(\mathbf x)\dd^d x.
\label{CF}
\end{equation}
We ask for the map $\mathbf M\in\mathcal M$ that minimizes the cost functional \eqref{CF}, i.e., such that $\mathsf E[\mathbf M;w]=\min_{\boldsymbol\mu\in\mathcal M}\mathsf{E}[\boldsymbol\mu;w]$. This problem is known in Measure Theory as the Monge transport problem \cite{Bogachev2012,Evans1997} and a lot of results have been obtained regarding the existence of the optimal map and its properties \cite{Villani2009}. One of the most interesting cases is the quadratic one, in which the cost is given by the convex function $w(\mathbf x,\mathbf y)=\|\mathbf x-\mathbf y\|^2$, and we have to minimize the functional
\begin{equation}
\mathsf{E}^{(2)}[\boldsymbol\mu]=\int_{{\mathsf T^d}}\|\mathbf x-\boldsymbol\mu(\mathbf x)\|^2\rho_1(\mathbf x)\dd^d x.
\label{squared}
\end{equation}
In the case of quadratic cost it can be proved that the optimal map can be expressed as the gradient of a certain function $\varphi$ \cite{Evans1997}, i.e. $\mathbf M(\mathbf x)=\grad\varphi(\mathbf x)$. Eq.~\eqref{h} can be than rewritten in terms of  $\varphi$, obtaining the so called \textit{Monge--Amp\`ere equation}
\begin{equation}
\rho_1(\mathbf x)=\rho_2(\grad\varphi(\mathbf x))\det\Hess\,\varphi(\mathbf x),
\end{equation}
where $\left(\Hess\,\varphi(\mathbf x)\right)_{i j}=\frac{\partial^2\varphi(\mathbf x)}{\partial x_i\partial x_j}$ is the Hessian matrix of $\varphi$.

Suppose now that $\rho_1(\mathbf x)=1+\delta\rho_1(\mathbf x)$ and $\rho_2(\mathbf x)=1+\delta\rho_2(\mathbf x)$ , being $\abs{\delta\rho_1(\mathbf x)}\ll 1$ and $\abs{\delta\rho_2(\mathbf x)}\ll 1$ $\forall\mathbf x\in{\mathsf T^d}$. We expect that, under these hypothesis, we can write $\mathbf M(\mathbf x)=\mathbf x+\mathbf m(\mathbf x)$ with $\norm{\mathbf m(\mathbf x)}\ll 1$ $\forall\mathbf x\in{\mathsf T^d}$: in the first order approximation, $\det\J_{\mathbf M}(\mathbf x)\approx 1+\dive\mathbf m(\mathbf x)$, so Eq.~\eqref{h} becomes:
\begin{equation}
\dive\mathbf m(\mathbf x)=\rho_1(\mathbf x)-\rho_2(\mathbf x)\defeq \delta\rho(\mathbf x).
\end{equation}
In particular, if $w(\mathbf x,\mathbf y)$ has the form as in Eq. \eqref{squared} we can introduce $\mathbf m(\mathbf x)=\grad\phi(\mathbf x)$, obtaining the simple Poisson equation
\begin{equation}
\Delta\phi=\delta\rho.
\label{poiss}
\end{equation}
Denoting by $\dhr_\nn\defeq\int_{\mathsf T^d}\delta\rho(\mathbf x)\e^{-2\pi i\mathbf n\cdot\mathbf x}\dd^dx$, in this case the total cost of the transport is given at the first order by 
\begin{equation}
\mathsf{E}^{(2)}[\mathbf M]\approx\int_{\mathsf T^d} \left[\grad\phi(\mathbf x)\right]^2\dd^d x=\sumd\frac{\abs{\dhr_\nn}^2}{4\pi^2\norm{\mathbf n}^2}.
\label{densenergy}
\end{equation}
Although the last equation has been derived under assumptions difficult to justify in the discrete and random version of the  Monge--Kantorovi\v{c} problem, that is in the $\ebmp$, we will see how Eq. \eqref{densenergy} retains its validity also in that case.
 
\subsection{The scaling ansatz}

Inspired by the previous considerations, we made an ansatz about the functional dependence of the optimal cost of the Euclidean bipartite matching problem with quadratic cost from the density of the two sets of points. The ansatz is simple, yet it is surprisingly predictive.

We denote with $\rho_\blue(\mathbf x)\equiv\frac{1}{N}\sum_{i=1}^N\delta\left(\mathbf x-\bb_i\right)\equiv\rho_1(\mathbf x)$  and with $\rho_\red(\mathbf x)\equiv\frac{1}{N}\sum_{i=1}^N\delta\left(\mathbf x-\rr_i\right)\equiv\rho_2(\mathbf x)$ the random densities in $[0,1]^d$ of the $N$  $\blue$-points and $\red$-points respectively. We suppose that periodic boundary conditions are imposed, so we work on the torus $\mathsf{T}^d$, as explained in the introduction.
Let us call $\delta\rho(\mathbf{x})\defeq \rho_{\blue}(\mathbf{x})-\rho_{\red}(\mathbf{x})$ the difference between the two densities and
\begin{equation}
\dhr_\nn\defeq\frac{1}{N}\sum_{i=1}^{N}\left(\e^{- 2\pi i \nn\cdot\bb_i}-\e^{-2\pi i \nn\cdot\rr_i}\right)\qquad\nn\in \Z^d,
\end{equation}
its Fourier modes. Following the hint given by the continuous problem, Eq. \eqref{densenergy}, we introduce the following functional
\begin{equation}
\E_{N}[\dhr]\defeq \sumd \frac{\abs{\dhr_\nn}^2}{4\pi^2\norm{\nn}^2}.
\label{ansatz}
\end{equation}
Our hypothesis is that the functional $\E_{N}[\dhr]$ at large $N$ captures the leading terms of the exact optimal cost $E_{N}^{(2)}[\pi^*;\{\mathbf r_i,\mathbf b_i\}]$, i.e. asymptotically $ E_{N}^{(2)}[\pi^*;\{\mathbf r_i,\mathbf b_i\}]\sim\E_{N}[\dhr]$, in the notation of Section \ref{empSection}. Note that we are using only \eqref{poiss} to evaluate the scaling of the optimal cost, without any reference to the optimality conditions itself. However, this is sufficient to reproduce the correct average behaviour: in fact, in the limit of validity of our linearisation of the Monge--Amp\`ere equation, the solution of \eqref{poiss} on the torus is unique and therefore determines automatically the optimal map. It can be shown by direct calculation  that $\overline{\abs{\dhr_\nn}^2}=\frac{2}{N}$ for each $\nn\neq\mathbf 0$. Therefore we have
\begin{equation}\label{meane}
\bnd\sim N^{2\over d}\overline{\E_N[\dhr]}=\sumd \frac{N^{2-d\over d}}{2\pi^2\norm{\mathbf n}^2}.
\end{equation}
For $d\geq 2$ the sum in the previous relation is divergent. However, by means of a proper regularisation of the sum, we can still extract useful informations on the scaling of $\beta_N$.  For $d=2$ Eq.~\eqref{meane} provides, after the regularization procedure, the leading scaling behaviour with the correct prefactor, whilst for $d>2$ the procedure gives the leading scaling and the prefactor of the subleading  behaviour. Sadly, in no case for $d>2$ can the coefficient, which we name $e^{(2)}_d$, of the leading term in the $\beta_N$ expansion be computed using our formalism.

\subsubsection{Case $d=1$}
\begin{figure}[h]\centering
\subfigure[\ Numerical data for $\frac{1}{N}\bn(1)$.]{
\includegraphics{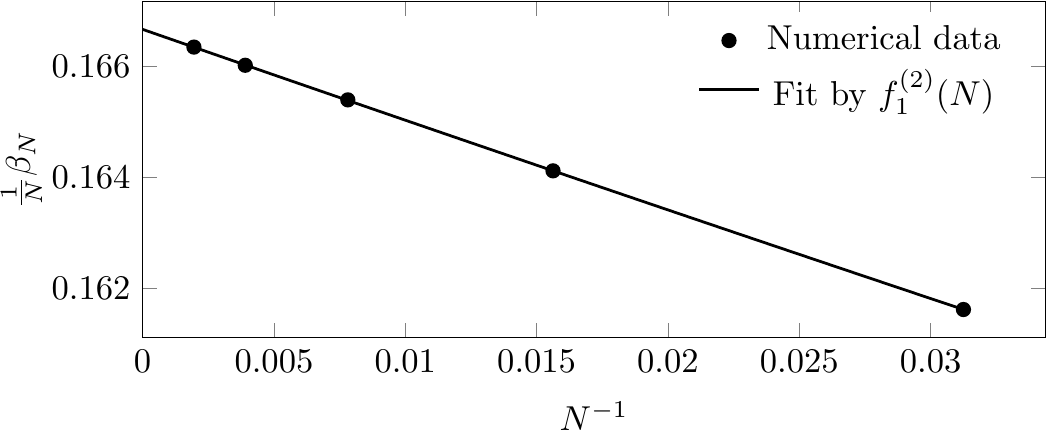}}
\hspace{1cm}\subfigure[\ Numerical results for $\bn(1)-\frac{N}{6}$.]{
\includegraphics{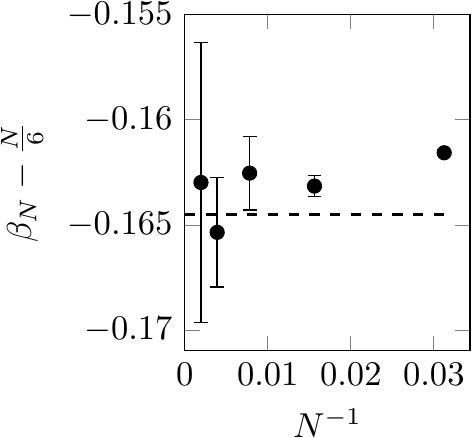}}
\caption{Numerical simulations for $d=1$; the fit was performed using the three parameters fit function $f^{(2)}_1(N)=\alpha^{(2)}_1+\frac{e_1^{(2)}}{N}+\frac{c}{N^2}$. }
\label{fig:d1}
\end{figure}
Let us consider the $\textsc{Ebmp}$ in dimension $d=1$. In low dimensions, that is for $d=1,2$, the optimal cost is no longer a subadditive quantity and the fluctuations of the density of points are dominant \cite{Houdayer1998}. Moreover, for $d=1$, the optimal cost  is also not self-averaging \cite{Houdayer1998}, while for $d=2$ this is still an open question. 

In one dimension the structure of the optimal matching is known \cite{Houdayer1998,Boniolo2014}: for any $p>1$ and using open boundary conditions, the optimal matching is the one associating the $k$-th blue point to the $k$-th red point, ordering the points from left to right along the line. This consideration leads to the prediction $\bn(1)=O(N^{\frac{p}{2}})$ for the leading behaviour with generic cost exponent $p$. For closed periodic boundary conditions, the case we are considering, a similar scenario holds: enumerating the red points and the blue points clockwise or anticlockwise, the optimal matching is given by a cyclic permutation, with offset to be determined by the optimality condition.

The one-dimensional case constitutes the simplest application of our formula, Eq.~\eqref{meane}, since this is the only where the sum is not divergent. We obtain straightforwardly
\begin{equation}
\bn(1)\sim\frac{N}{\pi^2}\sum_{n = 1}^{+\infty}\frac{1}{n^2}=\frac{N}{6}.
\label{E1}
\end{equation}
This is indeed the exact asymptotic behaviour  of $\beta_N$ \cite{Boniolo2014}, and it is a first validation of our very simple ansatz, Eq.~\eqref{ansatz}: we were able to catch both the anomalous scaling $O\left(N\right)$ and its correct prefactor.

We checked the validity of Eq.~\eqref{E1} averaging the  optimal cost of the \textsc{Ebmp} given by an exact algorithm \cite{lemon} for system sizes up to $N=2048$: here and in the following numerical simulations were performed using a C++ code and the open source \textsc{Lemon Graph Library} \cite{lemon}. We found the numerical data for $\frac{1}{N}\beta_N$ to be well approximated by a three parameters fitting function of the form $f^{(2)}_1(N)=\alpha^{(2)}_1+\frac{e^{(2)}_1}{N} + \frac{c}{N^2}$, where an additional correction of higher order is included. From a least square fit we obtained the coefficient $\alpha^{(2)}_1=0.166668(3)$, in perfect agreement with our analytical prediction (see Figure~\ref{fig:d1}).

Once we verified the validity of Eq.~\eqref{E1}, we used it to extrapolate the subleading coefficient $e^{(2)}_{1}$, fixing $\alpha^{(2)}_1=\frac{1}{6}$ and using the fitting function $f^{(2)}_1(N)$ with two free parameters  (see Figure~\ref{fig:d1} and Table \ref{tab:nu2}).

\subsubsection{Case $d=2$}
\begin{figure}[h]\centering
\subfigure[\ Numerical data for $\bn(2)$.]{
\includegraphics{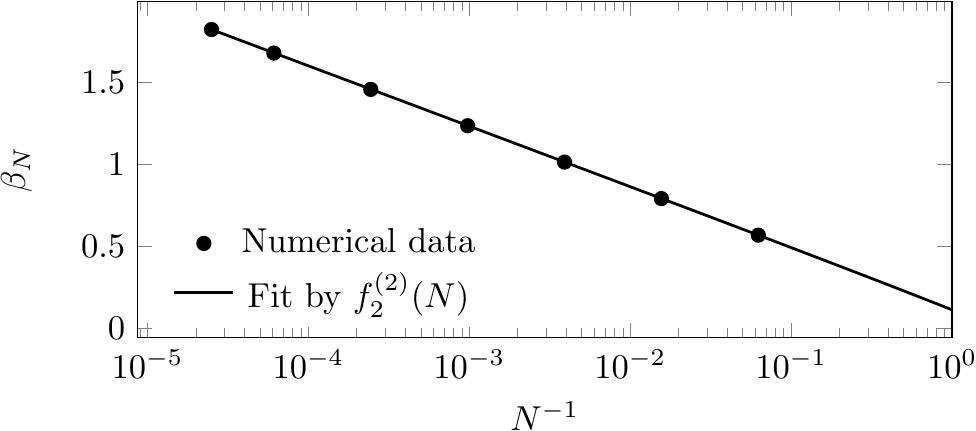}}
\hspace{1cm}\subfigure[\ Numerical results for $\bn(2)-\frac{\ln N}{2\pi}$.]{
\includegraphics{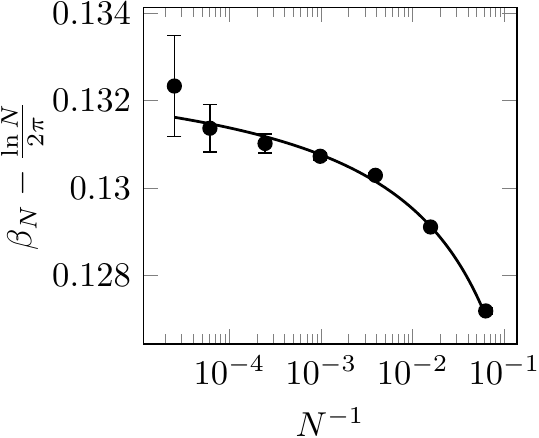}}
\caption{Numerical simulations for $d=2$. We fitted our numerical data for $\bn(2)$ using the function $f^{(2)}_2(N)=a\ln N+e_2^{(2)}+\frac{b}{\ln{N}}$. The $\frac{1}{\ln{N}}$ correction was suggested by the right hand plot.}
\label{fig:d2}
\end{figure}
As already stated in the Introduction, in dimension $2$ we have that $\beta_N=O\left(\ln N\right)$. In this paragraph we will show how to derive this scaling from our ansatz, Eq.~\eqref{meane}, and we will also obtain the corresponding prefactor. The sum appearing in Eq.~\eqref{meane} diverges in dimension $2$ and above, therefore we have to find a suitable regularization to give meaning to the expression. The regularization procedure presents some peculiarities at $d=2$ from which the anomalous scaling arises.

We choose a regularizing smooth function $F(x)$ such that $F(0)=1$  and $F(x)\to 0$ for $x\to+\infty$. The function has to decrease  rapidly enough to make the series $\sum_{\mathbf n\in\Z^2\setminus\{\mathbf 0\}}\frac{1}{\norm{\nn}^2}F\left(2\pi\norm{\mathbf n}\over 2\pi \ell^{-1}\right)$ converge: here we introduced a cut-off in momentum space, $2\pi \ell^{-1}$, where $\ell\equiv{1\over \sqrt{N}}$ is the characteristic length for the system, being $\ell$ of the order of the average distance between a blue point and the nearest red point. Clearly $2\pi \ell^{-1}\to +\infty$ for $N\to+\infty$. Finally, let us denote by $\mathcal N_d(r)=\abs{\{\mathbf x\in \Z^d\setminus\{\mathbf 0\}|\norm{\mathbf x}<r\}}$, the number of lattice points (excluded the origin) included in a ball of radius $r$ centred in the origin in dimension $d$. Then, for arbitrary $a\in(0,1)$, we can write
\begin{equation}
\begin{aligned}
\sumtwo\frac{1}{\norm{\mathbf n}^2}F\left(\frac{\norm{\mathbf n}}{\sqrt{N}}\right)&=\lim_{R\to\infty}
\int_a^{R}
\frac{1}{r^2}\, F\left(\frac{r}{\sqrt{N}}\right)\left[\frac{\partial\mathcal N_2(r)}{\partial r}-2\pi r\right]\dd r+2\pi\int_{a}^\infty F\left(\frac{r}{\sqrt{N}}\right)\frac{\dd r}{r}\\
&\approx\lim_{R\to\infty}
\int_a^{R}
\frac{1}{r^2}\left[\frac{\partial\mathcal N_2(r)}{\partial r}-2\pi r\right]\dd r+2\pi\int_{\frac{a}{\sqrt{N}}}^\infty \frac{F(r)}{r}\dd r,
\end{aligned}
\end{equation}
where we have isolated in the last term the singular part  of the series. The first integral in the right hand side is well behaved: indeed, $\int_a^{R}\frac{1}{r^2}\left[\frac{\partial\mathcal N_2(r)}{\partial r}-2\pi r\right]\dd r=\left.\frac{\mathcal N_2(r)-\pi r^2}{r^2}\right|_a^R+2\int_a^R\frac{\mathcal N_2(r)-\pi r^2}{r^3}$. Both the first and the second term are finite in the $R\to\infty$ limit due to the fact that \cite{Hardy1999} $\mathcal N_2(r)-\pi r^2\leq 1+2\sqrt{2}\pi r$. Therefore we have
\begin{equation}
\sumtwo\frac{1}{\norm{\mathbf n}^2}F\left(\frac{\norm{\mathbf n}}{\sqrt{N}}\right)\approx
\int_a^{+\infty}
\frac{2}{r^3}\,\left[\n_2(r)-\pi r^2\right]\dd r+\pi\log\frac{N}{a^2\e}+2\pi\int_{1}^\infty \frac{F(r)}{r}\dd r.
\end{equation}
Eq.~\eqref{meane} for the case $d=2$ can then be rewritten as
\begin{equation}
\bn(2)\sim\frac{\ln N}{2 \pi} +e_2^{(2)}.
\label{E2}
\end{equation}
where $e_2^{(2)}$ is some constant. To our knowledge the result $\lim_{N\to\infty}\frac{\bn(2)}{\ln N}=\frac{1}{2\pi}$ is new to the literature.

The validity of Eq.~\eqref{E2} has been confirmed by numerical simulation  with system sizes up to $N=4\cdot 10^4$. 
We found a three parameter function of the form $f^{(2)}_2(N)=a \ln N + e^{(2)}_2 + \frac{b}{\ln{N}}$ to be the best suited to fit the data for $\beta_N$. From a least square fit we obtain the coefficient $2 \pi a= 1.0004(6)$, in perfect agreement with our analytical prediction (see Figure~\ref{fig:d2}). Once verified the validity of Eq.~\eqref{E2}, we used it to extrapolate the subleading coefficient $e^{(2)}_{2}$, fixing $a=\frac{1}{2\pi}$ and fitting the other two parameters (see Figure~\ref{fig:d1} and Table~\ref{tab:nu2}).

\subsubsection{Case $d>2$} 
\begin{figure}[h]\centering
\subfigure[\ $d=3$.]{
\includegraphics{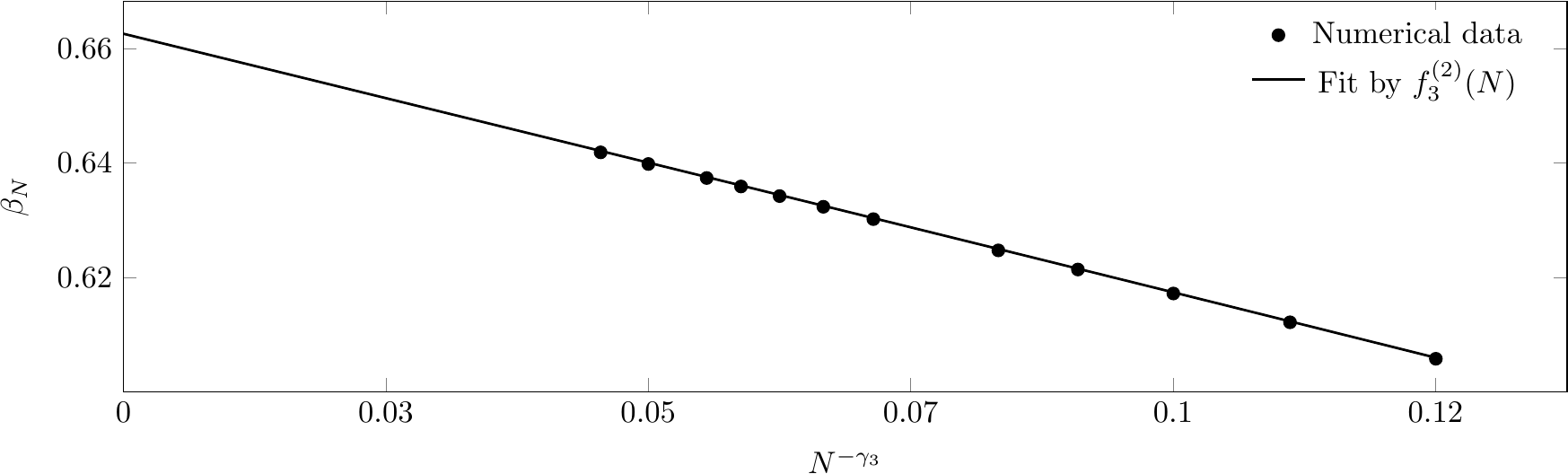}}\\
\subfigure[\ $d=4$.]{
\includegraphics{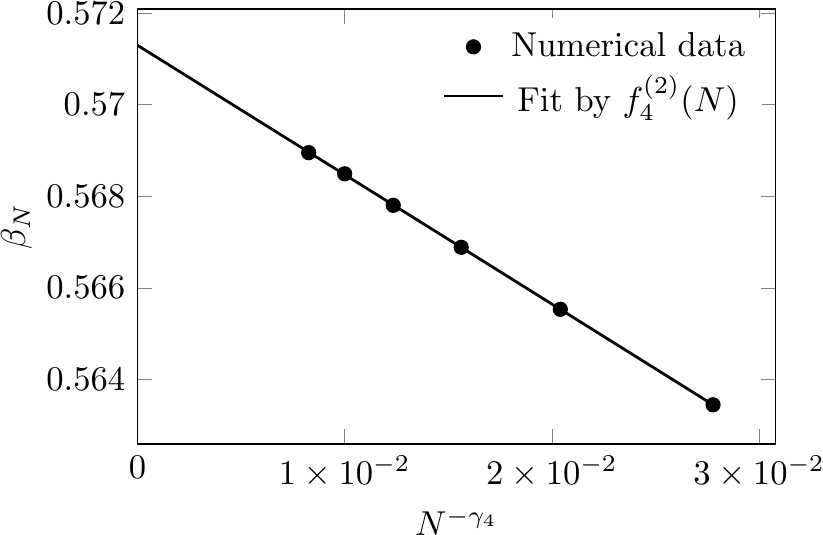}}
\subfigure[\ $d=5$.]{
\includegraphics{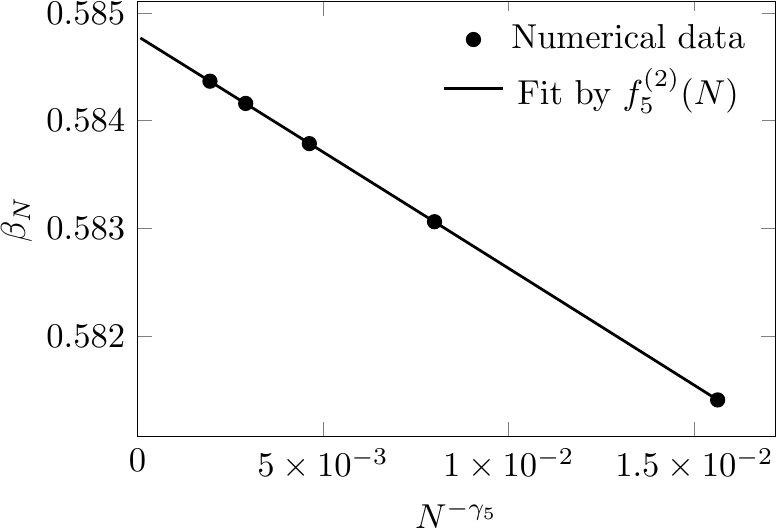}}
\caption{Numerical simulations for $d>2$. We fitted our numerical data for $\bn(d)$ using the function $f^{(2)}_d(N)=e^{(2)}_d+\alpha^{(2)}_d N^{-\gamma_d}+\frac{c}{N}$. The expected value for $\alpha^{(2)}_d$ was $\frac{1}{2\pi^2}\zeta_d(1)$.}
\label{fig:d345}
\end{figure}
\begin{table}
\begin{tabular}{c|ccccc}
\toprule
 & $d=1$ & $d=2$ & $d=3$   & $d=4$   & $d=5$\\
\hline
$\frac{1}{2\pi^2}\zeta_d(1)$ & $\frac{1}{6}$ & $-$ & $-0.45157\dots$ & $-0.28091\dots$ & $-0.21423\dots$ \\
$\alpha^{(2)}_d$ & $0.166668(3) $ & $-$&$-0.4489(16)$ & $-0.282(4)$ & $-0.2139(13)$ \\
$e^{(2)}_d$ & $-0.1645(13)$ & $0.1332(5)$ &  $0.66251(2)$  & $0.571284(6)$ & $0.584786(2)$
\end{tabular}
\caption{Results of numerical simulations for $p=2$. In the first line the analytical prediction for $\alpha^{(2)}_d$ for $d\neq 2$ is presented.}
\label{tab:nu2}
\end{table}
In dimensions greater than $2$ the average optimal cost has the leading scaling $E_N^{(2)}(d)=O(N^{-\frac{2}{d}})$ that one could expect from very simple arguments \cite{Mezard1985}, as already stated in the introduction. This is in fact the scaling obtained if each point is matched to one of its nearest points of different type, being their distance of order $O(N^{-\frac{1}{d}})$. Moreover it has been proven, using standard sub-additivity arguments, that $\beta_N$ is a self-averaging quantity and has a finite limit for $d>2$ \cite{Houdayer1998}. We will show that our ansatz Eq.~\eqref{ansatz} accounts for the subleading corrections to $\beta_N$. 

For $d\geq 2$ the series $\sumd\frac{1}{\norm{\mathbf n}^2}$ is  divergent. As in the previous paragraph, we use a sufficiently rapidly decaying function $F(x)$, with $\lim_{x\to\infty}F(x)=0$ and $\lim_{x\to 0}F(x)=1$, to regularize it. The characteristic length of the system is given by $\ell=\frac{1}{\sqrt[d]{N}}$. Denoting, as before, by $\mathcal N_d(r)$ the number of lattice points inside a sphere centred in the origin and of radius $r$, with the exclusion of the origin itself, we can write the following:
\begin{equation}
\begin{aligned}
\sumd\frac{1}{\norm{\mathbf n}^2}F\left(\frac{\norm{\mathbf n}}{\sqrt[d]{N}}\right)&=
\int_0^{+\infty}
\frac{1}{r^2}\, F\left(\frac{r}{\sqrt[d]{N}}\right)\left[\frac{\partial\n_d(r)}{\partial r}-S_d\,r^{d-1}\right]\dd r
+N^{d-2\over d}S_d\int_{0}^\infty F\left(r\right)r^{d-3}\dd r\\
&\approx
\int_0^{+\infty}
\frac{1}{r^2}\left[\frac{\partial\n_d(r)}{\partial r}-S_d\,r^{d-1}\right]\dd r
+N^{d-2\over d}S_d\int_{0}^\infty F\left(r\right)r^{d-3}\dd r,
\label{nfl}
\end{aligned}
\end{equation}
where $S_d=\frac{2\pi^{\frac{d}{2}}}{\Gamma\left(\frac{d}{2}\right)}$ is the unit sphere surface in $d$ dimensions. The last term in Eq.~\eqref{nfl} gives the leading-order contribution to $\beta_N$, but in our formalism it cannot be explicitly computed since it depends on the choice of the regularizing function $F(x)$. We name the other integral on the right hand side as
\begin{equation}
\Sigma_d\defeq\int_0^{+\infty}
\frac{1}{r^2}\left[\frac{\partial\n_d(r)}{\partial r}-S_d\, r^{d-1}\right]\dd r.
\label{sigmad}
\end{equation}
$\Sigma_d$ gives the first subleading correction to the leading scaling of $\beta_N$. Moreover, it can be shown (see appendix \ref{app:epstein}) that $\Sigma_d=\zeta_d(1)$, the analytic continuation to the point $s=1$ of the function
\begin{equation}
\zeta_d(s)\defeq\sumd\frac{1}{\norm{\mathbf n}^{2 s}} \qquad \text{for } \re s>\frac{d}{2}.
\end{equation}
The previous function is a particular case of a more general class of $\zeta$ functions, called \textit{Epstein zeta functions}. Therefore we are able to compute analytically the subleading behaviour of $\beta_N$ for $d>2$,
\begin{equation}\label{betad3}
\bn(d)\sim e_d^{(2)} +\frac{\zeta_d(1)}{2\pi^2N^{\gamma_d}},\qquad\gamma_d\equiv\frac{d-2}{d}.
\end{equation}
The expression for $\zeta_d(1)$ is given by Eq.~\eqref{analprol}, while the intensive costs $e_d^{(2)}$ have to be determined numerically. Note that for $d\to+\infty$ we recover the correct meanfield scaling behaviour already analyzed by  \citet{Houdayer1998} for the random assignment problem, i.e. $\gamma_d\to 1$. However, for finite $d$, the scaling behaviour can be very different from the mean field one.

We verified the validity of Eq.~\eqref{betad3} with numerical simulation on systems with sizes up to $N=10648$ in dimension $d=3$, $N=14641$ in dimension $d=4$ and $N=32768$ in dimension $d=5$. 
We used three parameter functions of the form $f^{(2)}_d(N)=e^{(2)}_d +  \frac{\alpha^{(2)}_d}{N^{\gamma_d}}+\frac{c}{N}$ to fit our data for $\bn(d)$. The scaling exponents $\gamma_d$ are readily confirmed to be the right ones (see Figure~\ref{fig:d345}) and the fitted coefficients $\alpha^{(2)}_d$ are in strong agreement with the analytical prediction $\frac{\zeta_d(1)}{2\pi^2}$ (Table~\ref{tab:nu2}). Then we fixed the  $\alpha^{(2)}_d=\frac{\zeta_d(1)}{2\pi^2}$ in $f^{(2)}_d(N)$ to extrapolate the extensive coefficients $e^{(2)}_d$, reported in Table~\ref{tab:nu2}.

\section{Results for generic $p$}
\begin{figure}[h]\centering
\subfigure[\ $d=3$.]{
\includegraphics{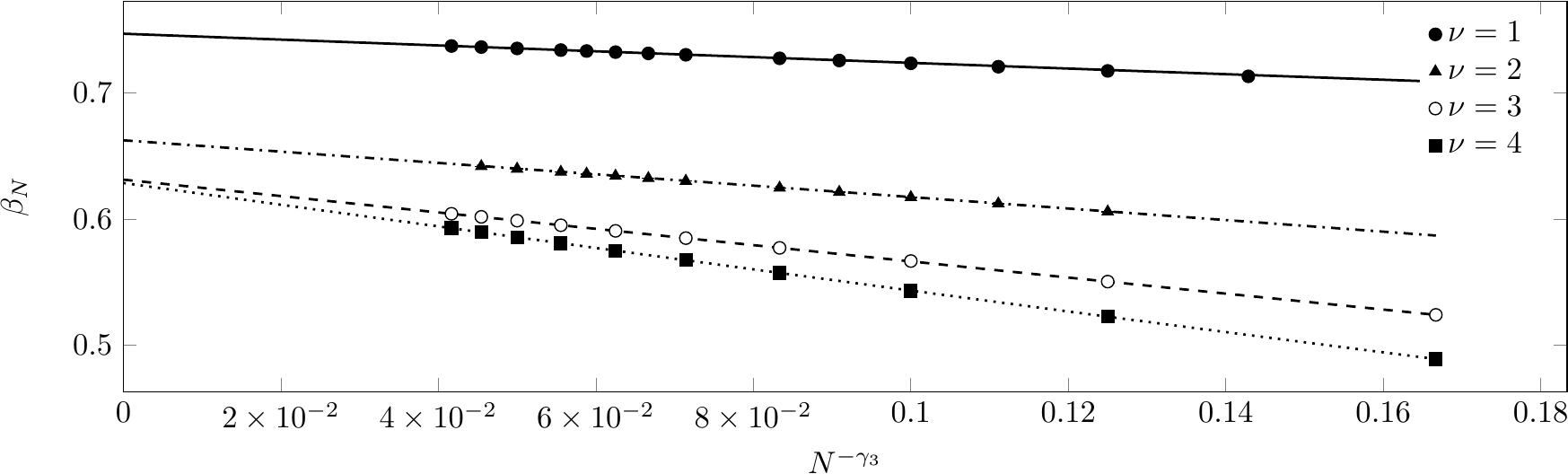}}\\
\subfigure[\ $d=4$.]{
\includegraphics{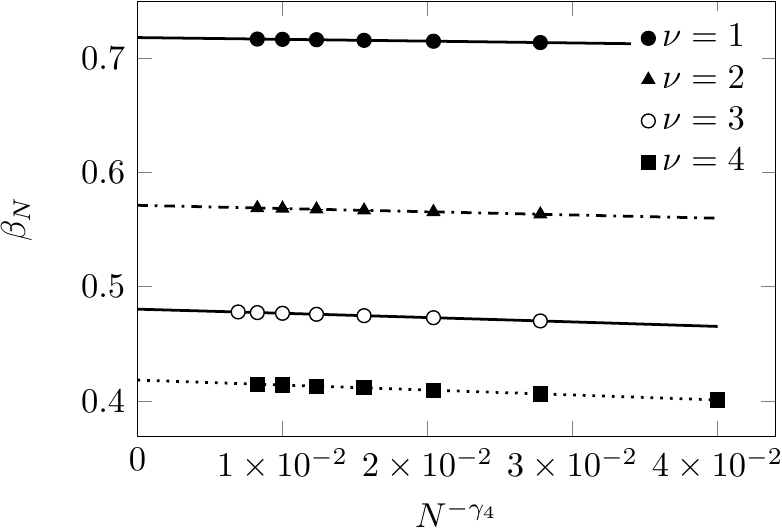}}
\subfigure[\ $d=5$.]{
\includegraphics{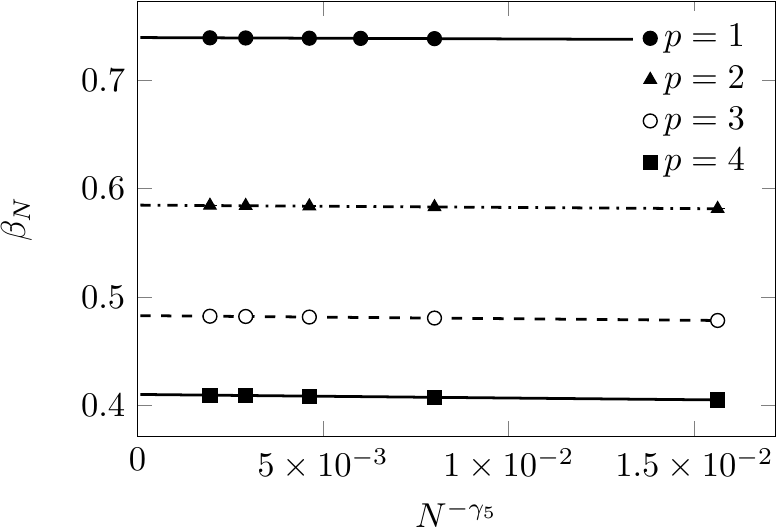}}
\caption{Numerical simulations for $d>2$. We fitted our numerical data for $\bnu(d)$ using the function $f_{d}^{(p)}(N)=e_d^{(p)}+\alpha_d^{(p)} N^{-\gamma_d}+\frac{c}{N}$. We plotted the results obtained for $p=1,3,4$. Fit results are presented in Table~\ref{tab:nu1}.}
\label{fig:nu1}
\end{figure}
\label{sec:generic}
\begin{table}
\begin{tabular}{l c c c c cc ccc}
\toprule
& \multicolumn{3}{c}{$p=1$}& \multicolumn{3}{c}{$p=3$}& \multicolumn{3}{c}{$p=4$}\\
  &  $d=3$   & $d=4$   & $d=5$&  $d=3$   & $d=4$   & $d=5$&  $d=3$   & $d=4$   & $d=5$\\
\hline
$\alpha^{(p)}_d$ & $-0.233(2)$ & $-0.152(17)$ & $-0.127(16)$ & $-0.652(3)$ & $-0.36(1)$ & $-0.28(4)$& $-0.863(3)$ & $-0.44(2)$ & $-0.34(1)$\\
$e^{(p)}_d$ & $0.7472(2)$  & $0.7181(2)$ & $0.73905(5)$ & $0.6313(2)$ & $0.4804(1)$ & $0.4830(1)$& $0.6286(2)$ & $0.4183(2)$ & $ 0.41043(4)$
\end{tabular}
\caption{Results of the fit of numerical data for $\bnu(d)$ by a function of the form $f^{(p)}_d(N)=e_d^{(p)}+\frac{\alpha_d^{(p)}}{N^{\gamma_d}}+ \frac{c}{N}$.}
\label{tab:nu1}
\end{table}

The asymptotic form we proposed for the average optimal cost $E_N^{(2)}(d)$ in the \textsc{Ebmp} with quadratic costs and periodic boundary conditions, Eq.~\eqref{ansatz}, could not be extended to cover the case of generic cost exponent $p$. Nonetheless, our numerical simulations give strong evidence that, for $d>2$ and any $p>0$, $\beta_N$  has  the asymptotic form 
\begin{equation}
\bnu(d)\sim e_d^{(p)}+\frac{\alpha_d^{(p)}}{N^{\gamma_d}},\qquad \gamma_d\defeq\frac{d-2}{d}.
\label{ansatzgenericnu}
\end{equation}
We thus find the same scaling exponent $\gamma_d$ analytically predicted only for the case $p=2$.
The nontrivial scaling exponent $\gamma_d$ differs from the mean-field exponent $\gamma_\infty=1$ of the random link matching problem \cite{Parisi2002} and of the Eucliden monopartite matching \cite{Houdayer1998}. The identification of the correct exponent $\gamma_d$ is crucial to the extrapolation of $e_d^{(p)}$ from numerical data. Since in the literature a mean-field-like scaling has been considered \cite{Houdayer1998}, we report in Table \ref{tab:nu1} the values of $e_d^{(p)}$ and $\alpha_d^{(p)}$ for different values of $p$. They were obtained fitting $\beta_N$ with functions of the form $f^{(p)}_d(N)=e_d^{(p)}+\frac{\alpha_d^{(p)}}{N^{\gamma_d}}+ \frac{c}{N}$. In Figure~\ref{fig:nu1} we plot $\beta_N^{(1)}$, $\beta_N^{(3)}$ and $\beta_N^{(4)}$ for $d=3,4,5$, along with the data already presented for $p=2$ for comparison. The scaling exponent $\gamma_d=\frac{d-2}{d}$ is confirmed by our simulations. Generalizing the case $p=2$, we therefore conjecture that the optimal cost as a function of the difference of the density of point $\delta\rho(\mathbf x)$, for $\abs{\delta\rho(\mathbf x)}\ll 1$ and in dimension $d>2$, can be approximated by
\begin{equation}
\E^{(p)}_{N}[\dhr]\defeq A^{(p)}_d \sumd \frac{\abs{\dhr_\nn}^2}{4\pi^2\norm{\nn}^2}.
\label{ansatznu}
\end{equation}
where  $A^{(p)}_d$ are unknown parameters. From last equation, the asymptotic form  Eq. \eqref{ansatzgenericnu} for $\beta_N$ can be derived, but at odds with the case $p=2$ where $A^{(2)}_d\equiv 1$ for any $d$, the lack of knowledge of the value of the parameters $A^{(p)}_d$ forbids the analytic prediction of the subleading coefficients $\alpha_d^{(p)}$. We also notice that Eq. \eqref{ansatznu} cannot be extended to dimensions $d=1,2$, since it is incompatible with the scaling scenario depicted in Eq. \eqref{scenario}.

The ansatz given in Eq. \eqref{ansatz}  for $p=2$ has been \textit{a posteriori} confirmed by the correct prediction of both the exponents $\gamma_d$ and the subleading coefficients $\alpha_d^{(2)}$. On the other hand, for generic $p$, the ansatz \eqref{ansatznu} is only supported by the fact that it gives the correct exponents $\gamma_d$, which is the reason itself why it was introduced. Therefore we tried to verify the internal consistence of Eq.~\eqref{ansatznu}. In fact after some algebraic manipulations and averaging over the disorder one can derive
\begin{equation}
A^{(p)}_d\sim\frac{\norm{\mathbf n}^2\pi N^2 }{2} \ \overline{\left(E^{(p)}_N[\pi^*;\{\mathbf r_i,\mathbf b_i\}]-E_N^{(p)}\right)\abs{\dhr_\nn}^2}.
\label{A}
\end{equation}
where we used the notation introduced in Eq.~\eqref{optcost}. We computed numerically the right-hand side of the previous equation for $d=3$ and sizes up to $N=10648$. The computation is numerically quite heavy since the density fluctuation are small. While the left-hand side of Eq.~\eqref{A} is independent of the mode $\nn$ in the large $N$ limit, we observed a finite size effect that seemed to be best accounted for by the scaling form
$A^{(p)}_3(\nn, N)\sim A^{(p)}_3 + b\frac{1}{\sqrt[3]{N}} + c\left(\frac{\norm{\mathbf n}}{\sqrt[3]{N}}\right)^2$. Using this three parameters function to fit our whole set of data ($\norm{\nn}<10$) at fixed $p$, we extrapolate the values of $A^{(p)}_3$ that we report in Table~\ref{tab:A}. In the same table we compare these extrapolations with the predictions steaming from Eq. \eqref{ansatznu}, that is $A^{(p)}_d=\frac{\alpha^{(p)}_d}{\alpha^{(2)}_d}=\alpha^{(p)}_d \frac{2\pi}{\zeta_d(1)}$, where the coefficients $\alpha^{(p)}_d$ where computed in Table \ref{tab:nu1}. The agreement between the two different sets of values is quantitatively and qualitatively good, even though we cannot definitively affirm the validity of Eq.~\eqref{ansatznu} due to the complexity and imprecision of the procedure utilized to test it. A more sound verification  would be to manually excite one of the modes through the addition a position dependent shift of the form  $\epsilon\,\nn \cos(2\pi \nn \cdot\mathbf{x})$ to each randomly generated point in one of the sets. One should then observe a linear response of the totat cost to the variation of $\epsilon$ as predicted by Eq. \eqref{ansatznu}.

\begin{table}
\begin{tabular}{c|cccc}
\toprule
  & $p=1$ & $p=2$   & $p=3$   & $p=4$\\
\hline
$A^{(p)}_3$ from Eq.~\eqref{ansatznu} & $0.516(5)$
 & $1$ & $1.44(1)
$ & $ 1.908(4)$ \\
$A^{(p)}_3$ from Eq.~\eqref{A} & $0.51(3)$ & $ 0.99(2) $ & $1.46(3)$	 & $1.96(2)$
\end{tabular}
\caption{Values of $A^{(p)}_d$ for $d=3$, extrapolated from Eqs. \eqref{ansatznu} and \eqref{A} respectively, as explained in the main text. The error in the second row are upscaled by a factor ten from those given by our fitting program (gnuplot), to assure the compatibility with the case $p=2$.}
\label{tab:A}
\end{table}

\section{Conclusions and perspectives}
In the present work we proposed a simple form for the asymptotic behaviour of the average optimal cost, $E^{(2)}_N(d)$, in the Euclidean bipartite matching problem with quadratic costs and periodic boundary conditions. This ansatz, Eq.~\eqref{ansatz}, contains no free parameters and leads to a strong set of predictions in every dimension, resumed in Eq. \eqref{contribution}. 
The rescaled cost $\bn(d)\defeq\frac{E^{(2)}_N}{N^{-\frac{2}{d}}}$ for low dimensions is a diverging quantity in the $\ebmp$, at odds with the monopartite case. We were able to prove that $\lim_{N\to\infty}\frac{\bn(1)}{N}=\frac{1}{6}$ and $\lim_{N\to\infty}\frac{\bn(2)}{\log N}=\frac{1}{2\pi}$ for $d=1$ and $d=2$ respectively.
Above the critical dimension of the system, $d=2$, the rescaled cost $\bn(d)$ has a finite limit which is inaccessible to our formalism. We were able though to determine analytically both the subleading scaling, $O(N^{-\gamma_d})$ with $\gamma_d=\frac{d-2}{d}$, and its prefactor $\frac{\zeta_d(1)}{2\pi^2}$. All the above claims are overwhelmingly supported by numerical simulations.

Finally, we provided numerical evidences that, in dimension $d>2$, the subleading scaling exponent $\gamma_d$ we predicted for the case of quadratic costs is the same for arbitrary cost exponent $p$. This led us to extend the ansatz proposed for quadratic costs, Eq. \eqref{ansatz}, to the general form Eq.~\eqref{ansatzgenericnu} for $d>2$. We tested numerically the validity of Eq.~\eqref{ansatzgenericnu}, obtaining good but not definitive results, therefore we proposed another numerical procedure that could give a firmer validation to the theory.

Although our scaling ansatz proved itself to be very powerful, as discussed above, a deeper theoretical investigation is required to derive analytically  the limit of $\beta^{(p)}(d)$ at large $N$, not computable in our framework. Moreover, a rigorous justification of our simple ansatz is desirable and could be inspired by the considerations we made in Section \ref{sec:monge} on the  Monge--Kantorovi\v{c} problem.
\subsection*{Acknowledgements}
The research leading to these results has received funding from the European Research Council under the European Union's Seventh Framework Programme (FP7/2007-2013) / ERC grant agreement No. 247328. S.C. and G.S. are grateful to Luigi Ambrosio for useful and stimulating discussions.

\appendix
\section{Evaluation of $\Sigma_d$}
\label{app:epstein} 
In this appendix we will show that, in dimension $d>2$, the function
\begin{equation}
\zeta_d(s)=\sumd\frac{1}{\norm{\nn}^{2 s}},
\label{zetads}
\end{equation}
a particular Epstein zeta function defined for $\re s > \frac{d}{2}$, analytically continued  to  the point $s=1$, takes the value of $\Sigma_d$  defined in Eq.~\eqref{sigmad}, i.e. $\zeta_d(1)=\Sigma_d$. Then we will derive an easily computable representation of the analytic continuation of $\zeta_d(s)$ which was already presented in the literature\citep{Ajtai1984,Contino2003}.

Let us fix $d>2$ and $a>0$. Then, for $\re s > \frac{d}{2}$, we can rewrite Eq.~\eqref{zetads} as
\begin{equation}
\begin{aligned}
\zeta_d(s)&=\lim_{R\to+\infty}\left[\sum_{\nn \in \Z^d\setminus \{0\}  }^{\norm{\nn} \leq R} \frac{1}{\norm{\nn}^{2s}}-S_d\int_a^R \frac{r^{d-1}}{r^{2s}}\dd r+S_d\int_a^R \frac{r^{d-1}}{r^{2s}}\dd r\right]\\
&=\lim_{R\to+\infty}\left[\sum_{\nn \in \Z^d\setminus \{0\}  }^{\norm{\nn} \leq R} \frac{1}{\norm{\nn}^{2s}}-S_d\int_a^R \frac{r^{d-1}}{r^{2s}}\dd r\right] +S_d\frac{a^{d-2s}}{2s-d}.
\end{aligned}
\end{equation}
Assuming that the limit in the last equation exists also for $\re s < \frac{d}{2}$, we have isolated the singular term. The analytic continuation of $\zeta_d(s)$ then reads
\begin{equation}
\zeta_d(s)=\lim_{R\to+\infty}\left[\sum_{\nn \in \Z^d\setminus \{0\}  }^{\norm{\nn} \leq R} \frac{1}{\norm{\nn}^{2s}}-S_d\int_0^R \frac{r^{d-1}}{r^{2s}}\dd r\right]  \qquad\text{for } \re s < \frac{d}{2},
\label{zetads2}
\end{equation}
where to limit $a\to 0$ has been taken. Note that, comparing the previous equation with Eq.~\eqref{sigmad}, $\zeta_d(1)\equiv\Sigma_d$. On the other hand, for $\re s > \frac{d}{2}$
\begin{equation}
\begin{aligned}
\zeta_d(s)&=\sumd\frac{1}{\Gamma(s)}\int_0^{+\infty}z^{s-1}\e^{-\norm{\nn}^2 z}\dd z
\\
&=\frac{\pi^s}{\Gamma(s)}\int_0^{+\infty}z^{s-1}\left(\Theta^d(z)-1\right)\dd z,
\end{aligned}
\end{equation}
where $\Theta(z)$ is defined by $\Theta(z)\defeq\sum_{n=-\infty}^{+\infty}\e^{-\pi n^2 z}\equiv\vartheta(0;i z)$, where $\vartheta(\tau;z)$ is the Jacobi $\theta$ function. Noticing that for $z\ll 1$ asymptotically we have $\Theta(z)\sim\frac{1}{\sqrt{z}}$, while  $\Theta(z)\sim 1+2\e^{-z}$ for $z\gg 1$, we can isolate the singular parts of $\zeta_d(s)$ writing
\begin{equation}
\zeta_d(s)=\frac{\pi^s}{\Gamma(s)}\left[\frac{2}{2s-d}-\frac{1}{s}+\int_1^{+\infty}z^{s-1}\left(\Theta^d(z)-1\right)\dd z+\int_0^{1}z^{s-1}\left(\Theta^d(z)-\frac{1}{z^{\frac{d}{2}}}\right)\dd z\right].
\label{zetadsfinal}
\end{equation}
The last expression can be readily continued to the region $\re s < \frac{d}{2}$. Using the property $\sqrt{t}\,\Theta(t)=\Theta(t^{-1})$, we can write
\begin{equation}
\Sigma_d=\zeta_d(1)=\pi\left[\frac{2}{2-d}-1+\int_1^{+\infty}\left(1+z^{\frac{d}{2}-2}\right)\left(\Theta^d(z)-1\right)\dd z\right].
\label{analprol}
\end{equation}

\bibliography{Biblio.bib}

\end{document}